\def\ra{\rangle}
\def\la{\langle}
\def\be{\begin{equation}}
\def\ee{\end{equation}}
\def\ba{\begin{array}}
\def\ea{\end{array}}

\documentclass[aps,pre,preprint,showpacs,amsmath,amssymb,amsfonts,preprintnumbers]
{revtex4}
\usepackage{epsfig,graphicx}

\begin{document}

\baselineskip=18pt \setcounter{page}{1} \centerline{\large\bf
On Estimation of Fully Entangled Fraction} \vspace{4ex}
\begin{center}
Rui-Juan Gu $^{1}$, Ming Li$^{2}$, Shao-Ming Fei$^{1}$ and Xianqing
Li-Jost$^{3,4}$

\vspace{2ex}

\begin{minipage}{5.5in}

\small $~^{1}$ {\small School of Mathematical Sciences, Capital
Normal University, 100048 Beijing}

\small $~^{2}$ {\small College of Mathematics and Computational
Science, China University of Petroleum, 257061 Dongying}

{\small $~^{3}$ Max-Planck-Institute for Mathematics in the
Sciences, 04103 Leipzig}

\small $~^{4}$ {\small Department of Mathematics, Hainan Normal
University, 571158 Haikou}

\end{minipage}
\end{center}

\begin{center}
\begin{minipage}{5in}
\vspace{2ex}
\centerline{\large Abstract}
\vspace{1ex}
We study the fully entangled fraction (FEF) of arbitrary mixed states. New upper bounds
of FEF are derived. These upper bounds make complements
on the estimation of the value of FEF. For weakly mixed quantum
states, an upper bound is shown to be very tight to the exact value of FEF.

\smallskip
PACS numbers: 03.67.-a, 02.20.Hj, 03.65.-w\vfill
\smallskip
\end{minipage}\end{center}
\bigskip

Quantum entanglement plays crucial roles in quantum information
processing such as quantum computation \cite{nielsen, di}, quantum
teleportation \cite{teleportation,teleportation1}, dense coding \cite{dense},
quantum cryptographic schemes \cite{schemes}, entanglement swapping
\cite{swapping} and remote states preparation (RSP)
\cite{RSP1,RSP2,RSP4}. For instance in terms of a
classical communication channel and a quantum resource (a nonlocal
entangled state like an EPR-pair of particles), the teleportation
protocol gives ways to transmit an unknown quantum state from a
sender to a receiver that are
spatially separated. When the sender and receiver share a maximally entangled pure state,
the state can be perfectly teleported.
However when the shared entangled state is an arbitrary mixed state $\rho$,
then the optimal fidelity of teleportation is given by \cite{teleportation1,yang},
\begin{eqnarray}
f(\rho)=\frac{d{\cal {F}}(\rho)}{d+1}+\frac{1}{d+1},
\end{eqnarray}
which solely depends on the fully entangled fraction (FEF) ${\cal
{F}}(\rho)$ of $\rho$. Detailed discussion about FEF can be found in
Ref. \cite{intr}.

In fact the quantity FEF plays essential roles in many other quantum
information processing such as dense coding, entanglement swapping
and quantum cryptography (Bell inequalities). Thus it is very
important to compute the FEF of general quantum states.
Unfortunately, precise formula of FEF has been only obtained for two
qubits systems \cite{grondalski}. For high dimensional systems, it
becomes quite difficult to derive an analytic formula for FEF. In
\cite{li} we have derived an upper bound of FEF to give an
estimation of the value of FEF. In this paper, we derive more tight
upper bounds for FEF. These bounds make complements on the
estimation of FEF.

Let ${\mathcal {H}}$ be a $d$-dimensional complex vector space with
computational basis $|i\ra$, $i=1,...,d$. The fully entangled
fraction of a density matrix $\rho\in{\mathcal {H}}\otimes{\mathcal
{H}}$ is defined by
\begin{eqnarray}\label{def}
{\mathcal {F}}(\rho)=\max_{\phi\in\epsilon}\la\phi|\rho |\phi\ra,
\end{eqnarray}
where $\epsilon$ denotes the set of $d \times d$-dimensional
maximally entangled pure states. (\ref{def}) can be also alternatively
expressed as
\begin{eqnarray}\label{def1}
{\mathcal {F}}(\rho)=\max_{U}\la\psi_{+}|(I\otimes U^{\dag})\rho
(I\otimes U)|\psi_{+}\ra,
\end{eqnarray}
where the maximization is taken over all unitary transformations $U$,
$|\psi_{+}\ra=\frac{1}{\sqrt{d}}\sum\limits_{i=1}^{d}|ii\ra$ is the
maximally entangled state and $I$ is the corresponding identity
matrix.

Let $h$ and $g$ be $d\times d$ matrices such that $h|j\ra=|(j+1)\mod
d\ra, $g$|j\ra=\omega^j|j\ra$, with
$\omega=exp\{\frac{-2i\pi}{d}\}$. We can introduce  $d^2$
linear-independent $d\times d$-matrices $U_{st}=h^{t}g^s$, which
satisfy \be U_{st}U_{s't'}=\omega^{st'-ts'}U_{s't'}U_{st}, {\rm
Tr}(U_{st})=d\delta_{s0}\delta_{t0}. \label{Mat} \ee
$\left\{U_{st}\right\}$ also satisfy the condition for {\it bases of
unitary operators} in the sense of \cite{Wer00}, i.e. \be
\left\{\begin{array}{l}
tr \left(U_{st}U^+_{s't'}\right)=d \delta_{tt'}\delta_{ss'},\\
U_{st}U_{st}^+=I.
\end{array}\right.
\label{Wer}
\ee
$\left\{U_{st}\right\}$ form a complete basis of $d\times d$-matrices,
namely, for any $d\times d$ matrix $W$, $W$ can be
expressed as \be\label{3e} W=\frac{1}{d}\sum_{s,t}tr
(U_{st}^+W)U_{st}. \ee

From $\{U_{st}\}$, we can introduce the generalized Bell-states,
\be\label{bellbas} |\Phi_{st}\ra=(I\otimes
U^*_{st})|\psi_+\ra=\frac{1}{\sqrt{d}}\sum_{i,j}(U_{st})^*_{ij}|ij\ra
,{\rm and }~|\Phi_{00}\ra=|\psi_{+}\ra, \ee $|\Phi_{st}\ra$ are all
maximally entangled states and form a complete orthogonal normalized
basis  of ${\mathcal {H}}\otimes{\mathcal {H}}$.

{\bf{Theorem 1:}} For any quantum state $\rho\in{\mathcal
{H}}\otimes{\mathcal {H}}$, the fully entangled fraction defined
in $(\ref{def})$ and $(\ref{def1})$ fulfills the following
inequality: \be\label{newfef} {\mathcal
{F}}(\rho)\leq\max_j\{\lambda_j\},\ee where $\lambda_j$s are the
eigenvalues of the real part of the matrix
$M=\left(
    \begin{array}{cc}
      T & iT \\
     -iT & T \\
    \end{array}
    \right)$,
$T$ is a $d^{2}\times d^{2}$ matrix with entries
$T_{n,m}=\la\Phi_n|\rho|\Phi_{m}\ra$ and $\Phi_j$ is the maximally
entangled basis states defined in $(\ref{bellbas})$.

{\bf{Proof:}} From $(\ref{3e})$, any $d\times d$ unitary matrix $U$
can be represented as $U=\sum_{k=1}^{d^2}z_{k}U_{k},$ where
$z_{k}=\frac{1}{d}{\rm Tr}(U_k^{\dag}U)$, $U_k$ are the unitary matrices
defined in (\ref{Mat}). Define
\be
x_l=\left\{\begin{array}{l}
{\rm Re}[z_l], 1\leq l\leq d^2;\\
{\rm Im}[z_l], d^{2}< l\leq 2d^2
\end{array}\right.~ {\rm{and \quad}}U^{'}_l=\left\{\begin{array}{l}
U_l, 1\leq l\leq d^2;\\
i*U_l, d^{2}< l\leq 2d^2.
\end{array}\right.
\ee
Then the unitary matrix $U$ can be rewritten as
$U=\sum_{k=1}^{2d^2}x_{k}U^{'}_{k}$. The necessary unitary
condition of $U$, ${\rm{Tr}}(UU^{\dag})=d$, requires that $\sum_k x_k^2=1$.
Set
\be\label{beinsert} F(\rho)\equiv\la\psi_{+}|(I\otimes
U^{\dag})\rho (I\otimes U)|\psi_{+}\ra=\sum_{m,n=1}^{2d^2}x_m x_n
M_{mn},\ee
where $M_{mn}$ is the entry of the matrix $M$ defined in theorem.
From the hermiticity of $\rho$ it is easily verified
that \be\label{hhh} M_{mn}^{*}=M_{nm}.\ee

To maximize $F(\rho)$ under constraints we get the following
equation
\be\frac{\partial}{\partial
x_k}\{F(\rho)+\lambda(\sum_l x_l^2-1)\}=0.\ee
Taking into account $(\ref{hhh})$ we obtain an eigenvalue equation,
\be\label{insert} \sum_{n=1}^{2d^2}{\rm{Re}}[M_{k,n}]x_n=-\lambda
x_k.\ee

Therefore \be {\mathcal {F}}(\rho)=\max_U F(\rho)
\leq\max_j\{\eta_j\},\ee where $\eta_j=-\lambda_j$ is the
corresponding eigenvalues of the real part of matrix $M$.
$\hfill\Box$

The upper bound derived in \cite{li} says that for any
$\rho\in{\mathcal {H}}\otimes{\mathcal {H}}$, the fully entangled
fraction ${\mathcal {F}}(\rho)$ satisfies
\begin{eqnarray}\label{oub}
{\mathcal {F}}(\rho)\leq
\frac{1}{d^{2}}+4||N^{T}(\rho)N(P_{+})||_{KF},
\end{eqnarray}
where $N(\rho)$ denotes the correlation matrix with
entries $n_{ij}(\rho)$ given in the expression of $\rho$
\begin{eqnarray}\label{rho}
\rho=\frac{1}{d^{2}}I\otimes
I+\frac{1}{d}\sum\limits_{i=1}^{d^{2}-1}r_{i}(\rho)\lambda_{i}\otimes
I+\frac{1}{d}\sum\limits_{j=1}^{d^{2}-1}s_{j}(\rho)I\otimes
\lambda_{j}+\sum\limits_{i,j=1}^{d^{2}-1}n_{ij}(\rho)\lambda_{i}\otimes
\lambda_{j},
\end{eqnarray}
$\lambda_{i}$, $i=1,...,d^2-1$, are the generators of the $SU(d)$
algebra with $Tr \{\lambda_{i}\lambda_{j}\}=2\delta_{ij}$,
$r_{i}(\rho)=\frac{1}{2}Tr\{\rho\lambda_{i}(1)\otimes I\}$,
$s_{j}(\rho)=\frac{1}{2}Tr\{\rho I\otimes \lambda_{j}(2)\}$,
$n_{ij}(\rho)=\frac{1}{4}Tr\{\rho \lambda_{i}(1)\otimes
\lambda_{j}(2)\}$, $P_{+}$ stands for the projection operator to
$|\psi_+\ra$, $N(P_{+})$ is similarly defined to $N(\rho)$, $N^{T}$
stands for the transpose of $N$, $||N||_{KF}=Tr\sqrt{NN^{\dag}}$ is
the Ky Fan norm of $N$. This upper bound was used to improve the
distillation protocol proposed in \cite{gdp}. Here we show that the
upper bound in $(\ref{newfef})$ is different from that in
$(\ref{oub})$ by an example.

{\bf{Example 1:}} We consider the bound entangled state \cite{ho232}
\be\label{exam}\rho(a)=\frac{1}{8a+1}\left(%
    \begin{array}{ccccccccc}
      a & 0 & 0 & 0 & a & 0 & 0 & 0 & a\\
      0 & a & 0 & 0 & 0 & 0 & 0 & 0 & 0\\
      0 & 0 & a & 0 & 0 & 0 & 0 & 0 & 0\\
      0 & 0 & 0 & a & 0 & 0 & 0 & 0 & 0\\
      a & 0 & 0 & 0 & a & 0 & 0 & 0 & a\\
      0 & 0 & 0 & 0 & 0 & a & 0 & 0 & 0\\
      0 & 0 & 0 & 0 & 0 & 0 & \frac{1+a}{2} & 0 & \frac{\sqrt{1-a^{2}}}{2}\\
      0 & 0 & 0 & 0 & 0 & 0 & 0 & a & 0\\
      a & 0 & 0 & 0 & a & 0 & \frac{\sqrt{1-a^{2}}}{2} & 0 & \frac{1+a}{2}\\
    \end{array}%
    \right).\ee
From Fig. 1 we see that
for $0\leq a<0.572$, the upper bound in $(\ref{newfef})$ is
larger than that in $(\ref{oub})$. But for $0.572<a<1$ the upper
bound in $(\ref{newfef})$ is always lower than that in
$(\ref{oub})$, i.e. the upper bound $(\ref{newfef})$ is tighter than $(\ref{oub})$ in this case.

\begin{figure}[h]
\begin{center}
\resizebox{10cm}{!}{\includegraphics{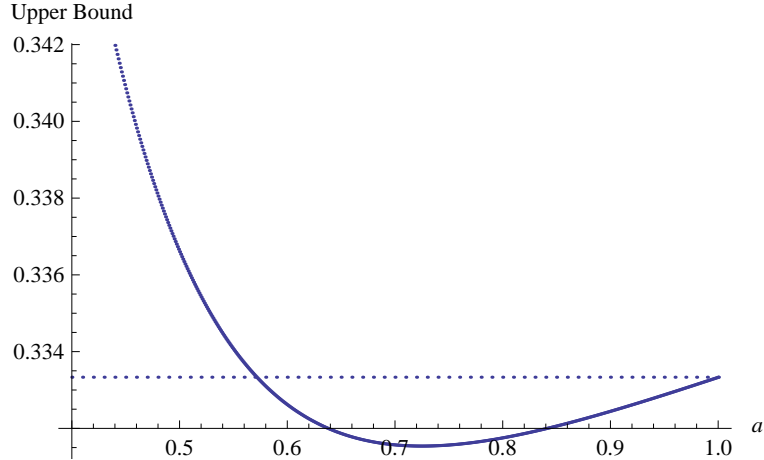}}
\end{center}
\caption{Upper bound of ${\mathcal {F}}(\rho)$ from
($\ref{newfef}$) (solid line) and upper bound from
$(\ref{oub})$ (dashed line). \label{fig1}}
\end{figure}

By using the operator norm, we have further

{\bf{Observation:}} For any $\rho\in{\mathcal {H}}\otimes{\mathcal
{H}}$, the fully entangled fraction ${\mathcal {F}}(\rho)$ satisfies
\be\label{22} {\mathcal {F}}(\rho)\leq \max_i\{\lambda_i\} \ee where
$\lambda_i$s are the eigenvalues of $\rho$.

 {\bf{Proof:}} For any quantum state $\rho$ and unitary $U$, we have
\begin{eqnarray}\label{18}
\la\psi_{+}|(I\otimes U^{\dag})\rho (I\otimes
U)|\psi_{+}\ra\leq\|\la\psi _{+}|\|\|(I\otimes
U^\dagger)\rho(I\otimes U)|\psi _{+}\ra\| \nonumber \\
\leq \|(I\otimes U^\dagger)\rho(I\otimes U)\|\||\psi_{+}\ra\|^{2}=
\|\rho\|\||\psi _{+}\ra\|^{2}=\|\rho\|,
\end{eqnarray}
where $\|\rho\|$ stands for the operator norm, $\|\rho\|=sup(\|\rho
 |x\ra\|:\||x\ra\|=1)$, $\||x\ra\|=\sqrt{\la x|x\ra}$.
We have used the Cauchy-Schwarz inequality to obtain the first
inequality. The second inequality is due to the basic property of
operator norm. The followed equality follows from the fact that
unitary transformation does not change the operator norm.

From \cite{conway} $\|\rho\|$ is an eigenvalue of $\rho$,
actually, it is the maximal eigenvalue of $\rho$, i.e.
$\|\rho\|=\max_i\{\lambda_i\}$ where $\lambda_i$s are the
eigenvalues of $\rho$, which ends the proof. $\hfill\Box$

This bound can give rise to further

{\bf Corollary: }Let $|\psi\ra=\sum_{ij}a_{ij}|ij\ra$ with
$|||\psi\ra||=1$ be the normalized eigenvector of $\rho$ with respect to the
maximal eigenvalue $\lambda_{max}$. If the matrix $A$ with elements
$A_{ij}=\sqrt{d}a_{ij}$ are unitary, the upper bound derived in
$(\ref{22})$ becomes the exact value of FEF.

{\bf{Proof:}} A simple computation shows that
\begin{eqnarray*}
{\mathcal{F}}(\rho)&&\leq\lambda_{max}=\la\psi|\rho|\psi\ra=\frac{1}{d}\sum_{ij,kl}
\sqrt{d}a_{ij}^{*}\la ij|\rho \sqrt{d}
a_{kl}|kl\ra \nonumber \\
&&=\sum_{i,k}\frac{1}{d}\la ii|I\otimes A^{\dag} \rho I\otimes
A|kk\ra =\la\psi_+|I\otimes A^{\dag} \rho I\otimes A|\psi_+\ra\leq
{\mathcal {F}}(\rho).\end{eqnarray*} Thus we have ${\mathcal
{F}}(\rho)=\lambda_{max}$. $\hfill\Box$

According to the corollary, we can find out when the upper bound
derived in theorem 2 becomes the exact value of FEF.

{\bf{Example 2:}} Consider the $3\otimes 3$ state \cite{prl1056}
\be\label{rr}\rho=\frac{2}{7}|\psi_{+}\rangle\langle\psi_{+}|+\frac{\alpha}{7}\sigma_{+}
+\frac{5-\alpha}{7}\sigma_{-},\ee where
$\sigma_{+}=\frac{1}{3}(|01\rangle\langle01|+|12\rangle\langle12|+|20\rangle\langle20|),
\sigma_{-}=\frac{1}{3}(|10\rangle\langle10|+|21\rangle\langle21|+|02\rangle\langle02|)$.
$\rho$ is entangled when $3<\alpha\leq5.$ The maximal eigenvalue of
$\rho$ is $\frac{2}{7}$, with the corresponding normalized
eigenvector
$\{\frac{1}{\sqrt{3}},0,0,0,\frac{1}{\sqrt{3}},0,0,0,\frac{1}{\sqrt{3}}\}$.
The matrix $A$ related to this eigenvector is just the $3\times 3$
identity matrix which is obviously unitary. Thus we have for the
state $(\ref{rr})$, ${\mathcal{F}}(\rho)=\frac{2}{7}$.

The following upper bound
of FEF gives a very tight estimation of FEF for weakly
mixed quantum states.

{\bf{Theorem 2:}} For any bipartite quantum state $\rho\in{\mathcal
{H}}\otimes{\mathcal {H}}$, the following inequality holds:
\be\label{fefm} {\mathcal {F}}(\rho)\leq
\frac{1}{d}({\rm{Tr}}\sqrt{\rho_A})^2,\ee where $\rho_A$ is the
reduced matrix of $\rho$.

{\bf{Proof:}} Note that the
FEF for pure state $|\psi\ra$ is given by \cite{gdp}
\be\label{fefp}{\mathcal
{F}}(|\psi\ra)=\frac{1}{d}({\rm{Tr}}\sqrt{\rho_A^{|\psi\ra}})^2,\ee
where $\rho_A^{|\psi\ra}$ is the reduced matrix of $|\psi\ra\la\psi|$.

For mixed state $\rho=\sum_ip_i \rho^i$, we have
\begin{eqnarray}\label{33}
{\mathcal {F}}(\rho)&=&\max_{U}\la\psi_{+}|(I\otimes U^{\dag})\rho
(I\otimes U)|\psi_{+}\ra\nonumber\\
&\leq&\sum_i p_i \max_U\la\psi_{+}|(I\otimes U^{\dag})\rho^i
(I\otimes U)|\psi_{+}\ra\nonumber\\
&=&\frac{1}{d}\sum_ip_i({\rm{Tr}}\sqrt{\rho_A^i})^2=\frac{1}{d}\sum_i({\rm{Tr}}\sqrt{p_i\rho_A^i})^2.
\end{eqnarray}
Let $\lambda_{ij}$ be the real and nonnegative eigenvalues of
the matrix $p_i\rho_A^i$. Recall that for any function
$F=\sum_i(\sum_jx_{ij}^2)^{\frac{1}{2}}$ subjected to the
constraints $z_j=\sum_ix_{ij}$ with $x_{ij}$ being real and
nonnegative, the inequality $\sum_jz_j^2\leq F^2$ holds.
It follows that
\begin{eqnarray}
{\mathcal
{F}}(\rho)\leq\frac{1}{d}\sum_i\left(\sum_j\sqrt{\lambda_{ij}}\right)^2
\leq\frac{1}{d}\left(\sum_j\sqrt{\sum_i\lambda_{ij}}\right)^2
=\frac{1}{d}({\rm{Tr}}\sqrt{\rho_A})^2,
\end{eqnarray}
which ends the proof. $\hfill\Box$

We now give an example to show that when the quantum state is weakly
mixed, theorem 3 will be a very good estimation for the FEF.

{\bf{Example 3:}} Consider the following $3\otimes3$ mixed state:
$\rho=\frac{1-p}{9}I_9+p|\psi\ra\la\psi|$, where
$|\psi\ra=\sqrt{\frac{x^2+2}{3}}\{x,0,0,0,1,0,0,0,1\}$ is a pure
state with one parameter $x$. To show the effectiveness of
(\ref{fefm}), we compare it with the single fraction of entanglement
$F_s=\la\psi_+|\rho|\psi_+\ra$. As seen from the Fig. 2, for weakly
mixed states (with larger parameter $p$), the bound provides
excellent estimation of the FEF.

\begin{figure}[tbp]
\begin{center}
\resizebox{16cm}{!}{\includegraphics{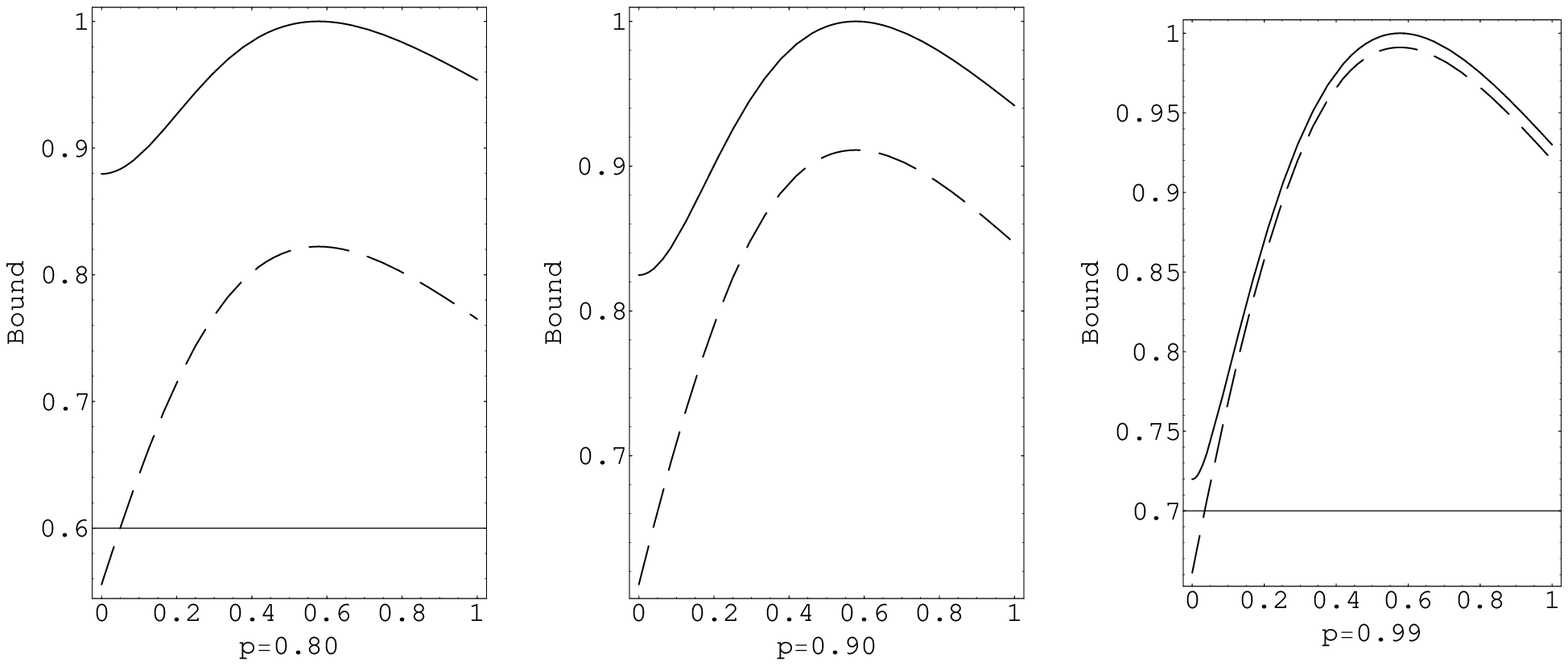}}
\end{center}
\caption{Upper bound of ${\mathcal {F}}(\rho)$ from
($\ref{fefm}$) (solid line) and single fraction of entanglement
$F_s$
\label{fig2}}
\end{figure}

\medskip
We have studied the fully entangled fraction that has tight
relations with many quantum information processing. New upper bounds
for FEF have been derived. They make complements on estimation of
the value of FEF. The conditions for the bounds to be exact or to be
more tight have been analyzed. These bounds provide a better
estimation of FEF and can be use in related information processing,
e.g. to detect the entanglement of the non-local source used in
quantum teleportation. Our results also give rise to good
estimations for the conditional min-entropy $H_{min}(A|B)$ and
$q_{corr(A|B)}$ according to relation in \cite{0807.1338}.

\bigskip
\noindent{\bf Acknowledgments}\,
This work is supported by NSFC 10675086, 10875081, 10871227,
KZ200810028013 and NKBRPC(2004CB318000).

\end{document}